\documentclass[sigconf]{acmart}
\AtBeginDocument{%
  }

\usepackage{microtype}
\usepackage{tabularx}
\usepackage{makecell}

\copyrightyear{2026}
\acmYear{2026}
\acmConference[GAS '26]{10th International Workshop on Games and Software Engineering }{April 12--18, 2026}{Rio de Janeiro, Brazil}
\acmBooktitle{10th International Workshop on Games and Software Engineering (GAS '26), April 12--18, 2026, Rio de Janeiro, Brazil}
\acmPrice{}
\acmDOI{10.1145/3786171.3788385}
\acmISBN{979-8-4007-2403-9/2026/04}

\setcopyright{none}

\newcommand{\datasetfinal}[0]{491 }
\newcommand{\datasetinit}[0]{10,548 }
\begin{document}

\title[Characterizing Open-Source Video Games from a Software Engineering Perspective]{Characterizing Open-Source Video Games from a \\Software Engineering Perspective}

\author{Henrique Mandelli Canella}
\email{hcanella@estudante.ufscar.br}
\orcid{0009-0000-7164-4864}
\affiliation{%
  \institution{Universidade Federal de São Carlos}
  \city{São Carlos}
  \state{São Paulo}
  \country{Brazil}
}

\author{Yohan Duarte}
\email{yohandp@estudante.ufscar.br}
\orcid{0000-0002-2700-711X}
\affiliation{%
  \institution{Universidade Federal de São Carlos}
  \city{São Carlos}
  \state{São Paulo}
  \country{Brazil}
}

\author{Cristiano Politowski}
\email{cristiano.politowski@ontariotechu.ca}
\orcid{0000-0002-0206-1056}
\affiliation{%
  \institution{Ontario Tech University}
  \city{Oshawa}
  \state{Ontario}
  \country{Canada}
}

\author{Vinicius Durelli}
\email{vinicius.durelli@ufscar.br}
\orcid{0000-0002-5768-1850}
\affiliation{%
  \institution{Universidade Federal de São Carlos}
  \city{São Carlos}
  \state{São Paulo}
  \country{Brazil}
}

\author{Andre Takeshi Endo}
\email{andreendo@ufscar.br}
\orcid{0000-0002-8737-1749}
\affiliation{%
  \institution{Universidade Federal de São Carlos}
  \city{São Carlos}
  \state{São Paulo}
  \country{Brazil}
}

\renewcommand{\shortauthors}{Canella et al.}

\begin{abstract}
Video games hold a prominent role in the entertainment industry, which constitutes a multibillion-dollar global market. 
Within this ecosystem, independent studios, small teams, and solo developers operate with constrained budgets while making their games available through platforms such as itch.io, where many projects are also available as open-source.
Where even if the project is fully released, other developers can still have access to the development process through the source code, helping other developers to develop their own games.
These open-source games also constitute a valuable corpus for empirical software engineering research, offering transparent access to development artifacts that are typically inaccessible in commercial projects.
This study characterizes open-source video games from a software engineering standpoint, 
focusing on their genres, popularity, 
repository activity, 
implementation characteristics, and testing practices.
To this end, we extracted metadata from 491 video games from itch.io, 
published as open-source projects in GitHub.   
Results show genre concentration in Puzzle, Action, and Platformer genres, conjectured to reflect lower development complexity and suitability for small teams. Available player ratings are high, suggesting community appreciation for open-source games. However, most repositories are dormant or minimally maintained, highlighting sustainability challenges. Projects predominantly adopt Godot or Unity engines, likely due to cost-effectiveness and open-source nature. Testing documentation is seldom reported, indicating underexplored quality assurance in open-source game development.
Our findings offer insights into the development of open-source video games, hypothesize reasons for observed patterns, and may contribute to advancing software engineering practices and 
tools designed for small teams and independent game developers by providing a baseline for testing and maintenance in open-source video games.
\end{abstract}

\ccsdesc[500]{Software and its engineering~Software organization and properties}
\ccsdesc[500]{Software and its engineering~Software libraries and repositories}

\keywords{Open-Source Video Games, Repository Mining, Software Testing}


\maketitle

\section{Introduction}

In 2023, the video game industry generated US\$93\,billion; console and PC titles led revenue.\footnote{\url{https://newzoo.com/resources/trend-reports/pc-console-gaming-report-2024}}
Analysts project US\$363\,billion by 2027 across PC, console, and mobile.\footnote{\url{https://www.statista.com/statistics/1344686/global-digital-gaming-revenue}}
This scale and release cadence shape how teams build and operate games.
Game development uses iterative cycles that cut across design, code, assets, and live operations; teams manage frame-rate budgets, platform compliance, engine and middleware updates, and frequent content drops \cite{novak2022}.
Secondary studies chart methods and evidence in Game Software Engineering \cite{engstrom_game_2018,Chueca2024,Literature_Software_Testing_Game_Industry}.

Despite the industry’s size, research underrepresents game-specific engineering, lacking comparable code-level evidence on key details like assets and engines. Prior work surveys developers and mines storefronts such as Steam\footnote{\url{https://store.steampowered.com}}
 and Itch.io\footnote{\url{https://itch.io}}, but closed repositories limit access to implementation details and histories. Open-source games can close this gap by exposing code, issues, pull requests, build scripts, tests, and release notes, enabling reproducible analyses of implementation choices, testing practices, dependency/engine usage, and evolution over time.

This paper characterizes open-source video games from a software engineering perspective. We analyze five components: (i) genres and player ratings; (ii) repository activity; (iii) implementation (engine use and asset composition); (iv) testing practices (automated and manual); and (v) relations among these components. We study \datasetfinal projects. Our pipeline validates repositories; collects GitHub metrics (commits, forks, contributors, issues); inspects source code to detect engines and count assets (images, audio, 3D models, shaders, fonts); and reviews tests and documentation to assess testing. 

Motivated by gaps in prior work, including surveys reporting low testing adoption without quantitative mining evidence \cite{kasurinen_what_2014, novak2022}, code-level insights extending beyond reproducibility \cite{Guidelines_Open_Source_Games}, and comprehensive characterizations linking genres, project activity, and quality practices \cite{truelove_well_2021},
we frame the study around five research questions: \textbf{RQ\(_1\)} What is the distribution of genres among open-source games? \textbf{RQ\(_{1.1}\)} How do genres differ in user ratings and rating volume? \textbf{RQ\(_2\)} How active are repositories in development and community engagement? \textbf{RQ\(_3\)} Which languages, engines/frameworks, and build/CI tools do projects use, and how large are they? \textbf{RQ\(_4\)} To what extent do projects document or adopt testing practices? \textbf{RQ\(_5\)} How is the presence of test files associated with repository activity and user-perceived quality? 

This paper contributes: (1) a metadata-only dataset of \datasetfinal open-source games that records repository activity, engine use, asset counts, and signals about testing and documentation; we manually verify entries for relevance and completeness; (2) an analysis of implementation and testing patterns in this corpus, showing that Godot and Unity appear most often, assets in 2D projects are mostly images, and only 7.33\% of repositories include automated tests; and (3) quantified relationships between project activity, technical characteristics, and testing practices that can guide follow-up studies and tooling.
By examining implications of genre and engine predominance (e.g., due to indie resource constraints) and exploring testing-practice associations with activity (but not ratings), this study snapshots the open-source game ecosystem, enabling research on sustainability and quality assurance improvements.

Our findings reveal that open-source games are concentrated in specific genres (Puzzle, Action, Platformer), maintain generally high player ratings when available, exhibit predominantly dormant repository activity, rely heavily on Godot and Unity engines, demonstrate simple asset compositions dominated by 2D images with limited 3D models, and show minimal adoption of automated testing (only 36 projects) with testing practices having no significant correlation with user-perceived quality.

The remainder of this paper is organized as follows. 
Section~\ref{sec:related_work} presents related work on game software engineering, emphasizing empirical studies of development practices, testing, and open-source ecosystems. 
Section~\ref{sec:study_setup} describes the study design, including data collection, filtering, and analysis procedures. Section~\ref{sec:threats_to_validity} discusses threats to validity. 
Section~\ref{sec:results} presents results structured around the research questions on genres, repository activity, implementation, testing practices, and associations. 
Section~\ref{sec:discussion} discusses implications within broader game development. 
Section~\ref{sec:concluding_remarks} concludes and outlines future work.

\section{Related Work}
\label{sec:related_work}

There are several literature reviews that aim to analyze the state-of-the-art in Game Software Engineering. 
For instance, 
\citet{engstrom_game_2018} explore the dual nature of video game development through a quantitative analysis of 228 studies, identifying a significant gap between Software Engineering and creative perspectives in existing research. Similarly, \citet{berg_marklund_what_2019} synthesize research on development practices, revealing that game development is characterized by an inherent inability to plan projects due to subjective requirements like ``fun'' and player experience. More specialized reviews include \citet{PolitowskiEPetrillo2021} analysis of video game testing practices and \citet{Aghyad2023} survey of game test automation methods, while \citet{Literature_Software_Testing_Game_Industry} provide a detailed classification of testing approaches and automation frameworks specifically for the game industry.

\begin{sloppypar}
Several studies employ surveys and interviews to understand game development practices from practitioners' perspectives. \citet{murphy-hill_cowboys_2014} explicitly contrast game development with traditional Software Engineering through interviews with 14 developers and a survey of 364 respondents, finding that game development prioritizes creativity and relies heavily on manual testing. \citet{kasurinen_what_2014} investigate testing priorities through 27 interviews, revealing that developers prioritize user experience and fun factor over technical reliability. More recently, \citet{Lambiase2025IEEETGAMES} examines Emotional Intelligence across job roles, identifying an ``emotional hierarchy'' where leaders bear disproportionate emotional labor. Complementary work by \citet{aleem_critical_2016} uses quantitative surveys to identify critical success factors, while \citet{borg_video_2020} study how practices adapt under extreme time pressure.
\end{sloppypar}

Our work aligns most closely with mining studies that analyze real-world game development artifacts. \citet{truelove_well_2021} conduct a large-scale analysis of 12,122 bug fixes from 723 Steam game updates, developing an expanded bug taxonomy and identifying `crash bugs' as most severe. Several studies focus on open-source ecosystems: \citet{Guidelines_Open_Source_Games} analyze 765 open-source game projects to propose reproducibility guidelines, evaluating if a game can be consistently built, set up, and run on local systems based on its corresponding open-source project, while \citet{Contribution_Guidelines_Software_testing} examine testing documentation in 200 non-games projects. In the mobile game domain, \citet{Issues_Industrial_Mobile_Games} analyze 91 real-world compatibility issues, and \citet{Playability_Issues_Mobile_Games} investigate playability through 20,346 user reviews. While these mining approaches provide empirical foundations that complement survey-based findings and offer scalable insights into actual development practices, they often focus on specific aspects such as bugs or reproducibility without linking to broader characteristics such as genres and repository activity. Our study addresses a comprehensive characterization of open-source games across multiple dimensions, including: 
genres, repository activity, implementation characteristics, and testing practices. This extends prior work by quantifying relations (e.g., testing and activity) and providing a dataset that enables further hypothesis-driven research, offering a better understanding of open-source game development that complements and extends previous researches.

\section{Study Setup}
\label{sec:study_setup}

\begin{figure*}[ht]
  \centering
  \includegraphics[width=1\textwidth]{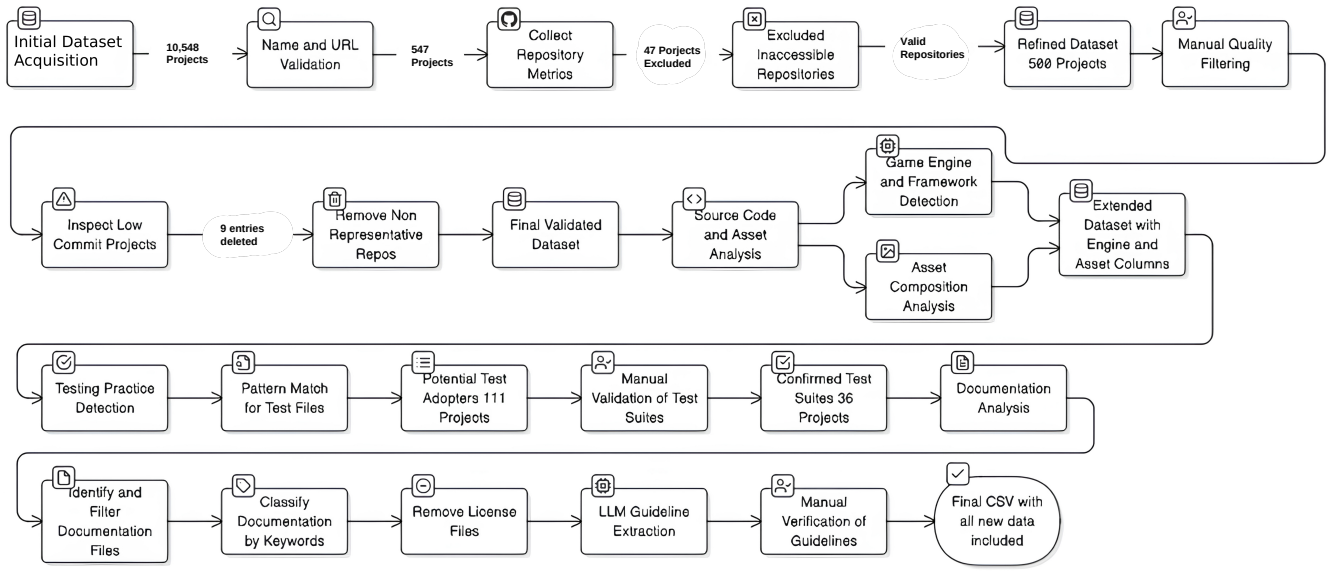}
    \caption{Flowchart of the study setup, starting with \datasetinit itch.io projects \cite{Guidelines_Open_Source_Games}, filtered to 491 through validation and quality checks. }
  \label{fig:setup}
\end{figure*}

\autoref{fig:setup} shows the multi-stage methodology we used to identify and analyze open-source game projects. The process combined automated data collection, data filtering, and manual validation of the dataset.

\paragraph{Initial Dataset Acquisition}
We used a dataset from \citet{Guidelines_Open_Source_Games}, which aggregates data from open-source games hosted on Itch.io platform. This initial dataset contained \datasetinit projects and included the following metadata: Name (name of the project), Rating, Author, Genre, Made with (tools and engines used through the development), Tags, Github\_links and Total ratings.

\paragraph{Name and URL Validation} 
We validated the mapping between project names and their GitHub URLs. Using fuzzy matching based on Levenshtein distance, we detected and corrected mismatches. This step produced a validated table of project–repository pairs and yielded a subset of 547 projects for analysis.

\paragraph{Collect Repository Metrics}
We used the GitHub API to collect per-repository metrics for the remaining projects: repository name, repository age, forks, stars, open issues, closed issues, total issues, average issue close time, and number of contributors. We merged these data with the initial dataset and excluded 10{,}048 entries without an accessible repository or retrievable data, yielding a corpus of 500 projects.

\paragraph{Manual Quality Filtering}
We added a manual screening step to keep the corpus focused on active game development rather than placeholders or misclassified projects. We ranked repositories by commit count and flagged those with 1–3 commits for review, since very short histories often indicate scaffolds or artifact drops. For each flagged repository, we inspected the default branch, commit messages, and file tree to verify the presence of versioned game source code (e.g., engine project files, code modules, build scripts) alongside assets. We retained repositories that showed a substantive initial import even if the total commits were small, to avoid penalizing new or migrated projects.

We excluded nine repositories that met at least one of the following conditions:
\begin{itemize}
    \item \textbf{Empty repositories:} Only an auto-generated \texttt{README.md} with no source files.
    \item \textbf{Archive-only code:} Code provided solely as an uploaded compressed file (e.g., \texttt{.zip}, \texttt{.rar}) rather than versioned sources.
    \item \textbf{Not a game:} Repositories that hosted engines, desktop tools, or standalone asset packs instead of a playable game.
\end{itemize}

For each exclusion, we recorded the rationale and removed the entry before analysis. This curation reduces label noise in downstream measures of activity, implementation characteristics, and testing indicators. After this step, the dataset contains \datasetfinal projects.



\paragraph{Engine and Framework Detection} We identified the primary engine or framework for each project by matching characteristic files and directories. We built a catalogue from the Itch.io engine list\footnote{\url{https://itch.io/game-development/engines/most-projects}} and mapped each technology to file patterns (e.g., Unity: \url{ProjectSettings/ProjectSettings.asset}, \texttt{*.unity}; Godot: \texttt{project.godot}; GameMaker: \texttt{*.yyp}; Construct: \texttt{*.c3p}). We scanned each repository for these signatures and assigned the most likely engine; ambiguous cases were marked as mixed and flagged for review.

\paragraph{Asset Inventory}
We quantified asset composition by counting files by extension across categories: images (\texttt{.png}, \texttt{.jpg}, \texttt{.jpeg}, \texttt{.bmp}, \texttt{.tga}, \texttt{.gif}, \texttt{.psd}, \texttt{.svg}); 3D models/meshes (\texttt{.fbx}, \texttt{.obj}, \texttt{.dae}, \texttt{.3ds}, \texttt{.blend}, \texttt{.gltf}, \texttt{.glb}); audio (\texttt{.mp3}, \texttt{.wav}, \texttt{.ogg}, \texttt{.flac}, \texttt{.aac}, \texttt{.m4a}); fonts (\texttt{.ttf}, \texttt{.otf}, \texttt{.fnt}); shaders (\texttt{.shader}, \texttt{.frag}, \texttt{.vert}, \texttt{.glsl}, \texttt{.hlsl}); and tilemaps/level data (\texttt{.tmx}, \texttt{.tsx}). 

\paragraph{Testing Practice Detection}
We screened repositories for candidate automated tests by matching test-related keywords in file and directory names (e.g., \texttt{test}, \texttt{tests}, \texttt{\_\_tests\_\_}, \texttt{spec}, \texttt{unittest}). This pass flagged 111 projects as potential adopters. We then manually verified the flagged set by inspecting source files and configuration (framework imports, assertion usage, test runners, and CI scripts) to separate true test suites from false positives (e.g., sample code or assets containing the word ``test''). This verification confirmed 36 projects with automated tests.

\paragraph{Documentation Analysis}
We identified documentation by scanning Markdown files whose paths contained \texttt{contributing}, \texttt{readme}, \texttt{development}, or \texttt{testing}, following \citet{Contribution_Guidelines_Software_testing}. We then applied a keyword filter\footnote{Keywords: bug, bugs, debug, debugged, debugging, defect, defects, error, fail, failure, flaws, flaw, issue, issues, report, reports, test, tested, testes, testing, tests.} to keep testing-related content and removed license files, yielding 180 candidates. Next, we used a local LLM (gpt-oss:20b)\footnote{\url{https://ollama.com/library/gpt-oss:20b}} to extract testing guidelines with the prompt: \emph{``Analyze the following text. If it contains ANY guidelines or instructions related to testing, bugs, defects, flaws, or reporting, summarize it in a few sentences, without topics and itemized. If there is nothing relevant, return ONLY: NO. Do not include or add extra text.''} (temperature 0.3). We then asked the model to assign each summary to one or more categories: Bug Related Activity, Testing Practices, Troubleshooting Practices, and Contribution Practices. We manually verified the guideline summaries on a sample and the category assignments on the full set. Finally, we consolidated artifacts and merged all tables into a single file, producing the final dataset of \datasetfinal items. Procedures and data are available at \url{https://zenodo.org/records/17451086}.

\section{Threats to Validity}
\label{sec:threats_to_validity}

The findings derive from open-source games on itch.io, limiting generalization to closed-source or other platforms. However, itch.io's large, diverse indie/open-source set (over 18,000 games \cite{Guidelines_Open_Source_Games}) suits our focus on resource-constrained empirical software engineering, offering underrepresented code-level artifacts compared to commercial sites like Steam. To counter selection bias from the third-party dataset \cite{Guidelines_Open_Source_Games}, which targeted itch.io open-source games for representativeness, we acknowledge potential metadata incompleteness. Automated threats (e.g., engine detection, asset counting) were minimized via manual validation on 20 random projects, confirming accuracy. Manual steps, like quality filtering and test verification, involved two independent researchers reviewing flagged repositories to ensure consistency and reduce bias.

\section{Results Analysis}
\label{sec:results}

\subsection{RQ$_1$:~What is the distribution of genres among open-source games?}

We measured genre assignments in the dataset. Table~\ref{tab:genres} reports counts by genre: \emph{Puzzle} (103), \emph{Action} (97), and \emph{Platformer} (69) account for the largest share; \emph{Adventure} (44), \emph{Shooter} (38), and \emph{Simulation} (38) follow. We labeled 56 games as \emph{Uncategorized} due to missing genre metadata. Because projects can list multiple genres, totals exceed the number of games.

\begin{table}[!htb]
    \caption{Distribution of open-source video games by genre.}
    \label{tab:genres}
    \centering
    \small
    \begin{tabular}{|l|c|}
        \hline
        \multicolumn{1}{|c|}{\textbf{Genre}} & \multicolumn{1}{c|}{\textbf{Number of Games}} \\
        \hline
        \hline
        \footnotesize Puzzle              & \small 103 \\
        \hline
        \footnotesize Action              & \small 97  \\
        \hline
        \footnotesize Platformer          & \small 69  \\
        \hline
        \footnotesize Uncategorized       & \small 56  \\
        \hline
        \footnotesize Adventure           & \small 44  \\
        \hline
        \footnotesize Shooter             & \small 38  \\
        \hline
        \footnotesize Simulation          & \small 38  \\
        \hline
        \footnotesize Strategy            & \small 27  \\
        \hline
        \footnotesize Interactive Fiction & \small 22  \\
        \hline
        \footnotesize Role Playing        & \small 18  \\
        \hline
        \footnotesize Survival            & \small 17  \\
        \hline
        \footnotesize Educational         & \small 14  \\
        \hline
        \footnotesize Visual Novel        & \small 14  \\
        \hline
        \footnotesize Racing              & \small 14  \\
        \hline
        \footnotesize Card Game           & \small 7   \\
        \hline
        \footnotesize Fighting            & \small 6   \\
        \hline
        \footnotesize Rhythm              & \small 5   \\
        \hline
        \footnotesize Sports              & \small 4   \\
        \hline
        \footnotesize Exploration         & \small 1   \\
        \hline
        \footnotesize Fantasy             & \small 1   \\
        \hline
        \footnotesize Medieval            & \small 1   \\
        \hline
        \footnotesize Open World          & \small 1   \\
        \hline
        \footnotesize Roguelike           & \small 1   \\
        \hline
        \footnotesize Roguelite           & \small 1   \\
        \hline
        \footnotesize Story Rich          & \small 1   \\
        \hline
        \footnotesize Text based          & \small 1   \\
        \hline
        \footnotesize Turn-based          & \small 1   \\
        \hline
        \footnotesize Life Simulation     & \small 1   \\
        \hline
        \footnotesize Sandbox             & \small 1   \\
        \hline
    \end{tabular}
\end{table}

\begin{figure*}[!htb]
  \centering
    \includegraphics[width=\textwidth]{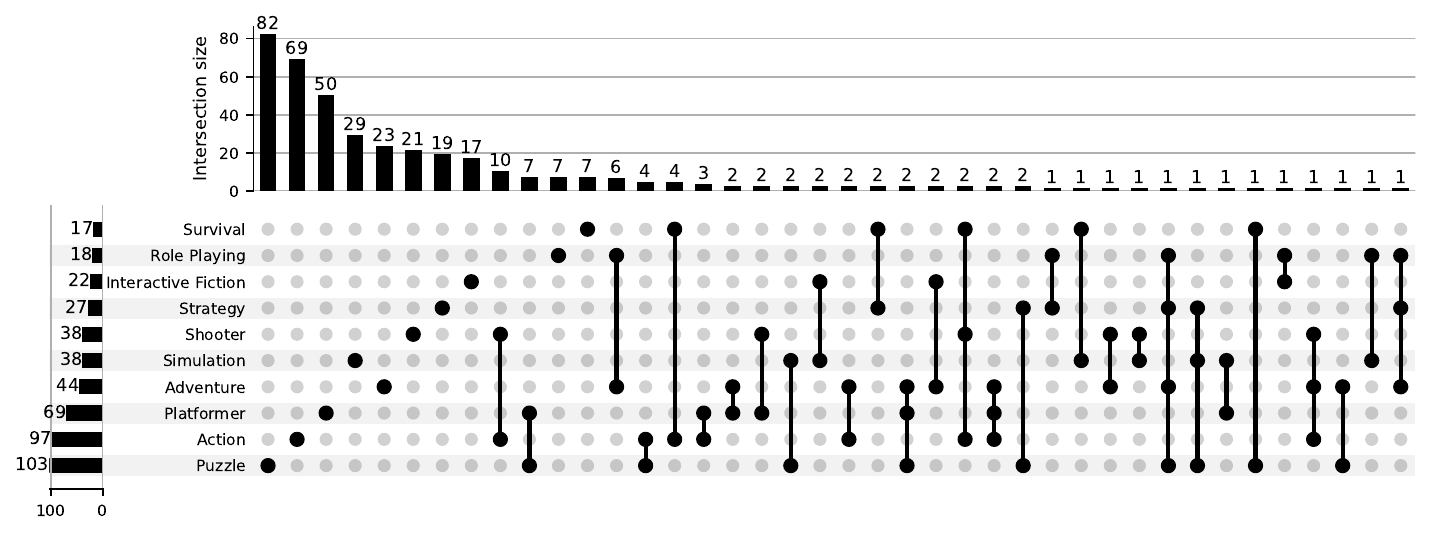}
    \caption{UpSet plot showing co-occurrence of top genres. Horizontal bars indicate genre set sizes; vertical bars show intersection sizes. Filled/empty circles denote included/excluded genres.}
  \label{fig:upset_top10_genres}
\end{figure*}

As the UpSet plot in Figure~\ref{fig:upset_top10_genres} shows \cite{upset_plot_ref}, most projects map to a single genre. A smaller group combines two or three genres; only one title (\texttt{Ben Was Assimilated}) spans four—\emph{Role Playing}, \emph{Strategy}, \emph{Adventure}, and \emph{Puzzle}. The largest multi-genre intersections involve \emph{Adventure}, \emph{Simulation}, and \emph{Shooter}. Single-genre labeling dominates, but cross-genre combinations appear regularly in the corpus.

\noindent
\textbf{Key insights for RQ$_1$:} 
Genres skew toward \emph{Puzzle}, \emph{Action}, and \emph{Platformer}. This concentration may stem from their relative simplicity in design and implementation, making them more feasible for solo or small-team developers with limited resources. Most projects carry a single genre label; a smaller share combines two or three, and only one spans four. Multi-genre intersections appear most often among \emph{Adventure}, \emph{Simulation}, and \emph{Shooter}, possibly indicating hybrid designs that leverage overlapping mechanics to enhance engagement without excessive complexity. Because games can hold multiple labels, totals exceed the number of projects; treat genre as a multi-label attribute in downstream analyses.

\subsubsection{\textbf{RQ$_{1.1}$: How do genres differ in terms of user ratings and rating volume?}}

We examined how genres differ in user ratings and rating volume. Of the 491 games, 289 lack ratings (58.9\%), indicating that player feedback is not consistently available; this absence may reflect early development status or limited visibility and engagement. Table~\ref{tab:genre_ratings} reports mean ratings and dispersion by genre on a 5-point scale. Most genres show means above 4.0. The highest means occur in \emph{Sports} (5.0) and \emph{Fighting} (4.75), each based on two titles, so small samples likely inflate these values. Among genres with at least 20 titles, \emph{Platformer} (4.72), \emph{Adventure} (4.61), and \emph{Action} (4.56) rank highest. \emph{Strategy} (4.39) and \emph{Survival} (4.18) sit lower but remain within the upper range of the scale. Standard deviations are typically below 0.7, indicating within-genre rating consistency. Overall, when ratings exist, evaluations are favorable across genres, but sparse coverage limits conclusions for many projects.

\begin{table}[!htb]
    \caption{Average rating and dispersion by genre.}
    \label{tab:genre_ratings}
    \centering
    \small
    \begin{tabular}{|l|c|c|c|c|c|}
        \hline
        \multicolumn{1}{|c|}{\textbf{Genre}} & 
        \multicolumn{1}{c|}{\makecell{\textbf{Number}\\\textbf{of Games}}} & 
        \multicolumn{1}{c|}{\textbf{Mean}} & 
        \multicolumn{1}{c|}{\textbf{SD}$^\ddag$} & 
        \multicolumn{1}{c|}{\textbf{Min}} & 
        \multicolumn{1}{c|}{\textbf{Max}} \\
        \hline
        \hline
        \footnotesize Action              & \small 40 & \small 4.56 & \small 0.62 & \small 3.0 & \small 5.0 \\
        \hline
        \footnotesize Puzzle              & \small 45 & \small 4.50 & \small 0.51 & \small 3.0 & \small 5.0 \\
        \hline
        \footnotesize Simulation          & \small 26 & \small 4.41 & \small 0.69 & \small 2.5 & \small 5.0 \\
        \hline
        \footnotesize Platformer          & \small 22 & \small 4.72 & \small 0.32 & \small 4.0 & \small 5.0 \\
        \hline
        \footnotesize Adventure           & \small 19 & \small 4.61 & \small 0.51 & \small 3.0 & \small 5.0 \\
        \hline
        \footnotesize Interactive Fiction & \small 16 & \small 4.55 & \small 0.49 & \small 3.0 & \small 4.9 \\
        \hline
        \footnotesize Shooter             & \small 14 & \small 4.57 & \small 0.60 & \small 3.0 & \small 5.0 \\
        \hline
        \footnotesize Strategy            & \small 14 & \small 4.39 & \small 0.68 & \small 3.0 & \small 5.0 \\
        \hline
        \footnotesize Role Playing        & \small 10 & \small 4.45 & \small 0.65 & \small 3.0 & \small 5.0 \\
        \hline
        \footnotesize Visual Novel        & \small 9  & \small 4.67 & \small 0.32 & \small 4.0 & \small 5.0 \\
        \hline
        \footnotesize Educational         & \small 7  & \small 4.60 & \small 0.45 & \small 4.0 & \small 5.0 \\
        \hline
        \footnotesize Survival            & \small 6  & \small 4.18 & \small 0.65 & \small 3.0 & \small 4.9 \\
        \hline
        \footnotesize Racing              & \small 4  & \small 4.60 & \small 0.49 & \small 4.0 & \small 5.0 \\
        \hline
        \footnotesize Rhythm              & \small 3  & \small 4.67 & \small 0.15 & \small 4.5 & \small 4.8 \\
        \hline
        \footnotesize Card Game           & \small 3  & \small 4.33 & \small 0.40 & \small 3.9 & \small 4.7 \\
        \hline
        \footnotesize Sports              & \small 2  & \small 5.00 & \small 0.00 & \small 5.0 & \small 5.0 \\
        \hline
        \footnotesize Fighting            & \small 2  & \small 4.75 & \small 0.35 & \small 4.5 & \small 5.0 \\
        \hline
    \end{tabular}
    \vspace{0.5em}
    \begin{flushleft}
        \footnotesize \phantom{\textit{Note.}} $^\ddag$SD stands for standard deviation.
    \end{flushleft}
\end{table}

\noindent
\textbf{Key insights for RQ$_{1.1}$:} 
Average player ratings are high across genres (Table~\ref{tab:genre_ratings}). Standard deviations are low for most genres, indicating stable evaluations. Overall reception is favorable regardless of genre, but coverage is limited: 58.9\% of projects lack ratings. This sparsity may reflect low visibility on platforms like itch.io, while high ratings could suggest that open-source video games appeal to niche, supportive communities that value innovation over polish.

\subsection{RQ$_2$: How active are the GitHub repositories of open-source video game projects in terms of development and community engagement?}

We measured repository activity across \datasetfinal projects (Table~\ref{tab:repo_activity}). Activity is right-skewed: the mean commit count is 333.38, yet 79 projects have $\le$5 commits, while the maximum reaches 65{,}860 commits. Recent development also skews toward low activity: the mean is 5.04 commits in the last six months, but the maximum is 954, indicating a small set of actively maintained projects alongside many dormant ones.

Community attention concentrates on a few repositories. Stars average 74.35 (maximum 11{,}963), and forks average 11.24 (maximum 1{,}966). Contributor counts show a similar pattern, with a mean of 4.22 and a maximum of 698.

Issue tracking is often absent. For more than 300 projects, total, open, and closed issues all equal zero, suggesting no recorded issues or no use of the tracker. Among projects that do record issues, the average close time is 7.58 days, while the maximum is 1{,}099 days, showing wide variation in maintenance throughput.

\begin{table}[!htb]
    \caption{Repository activity metrics.}
    \label{tab:repo_activity}
    \centering
    \footnotesize
    \begin{tabular}{|l|r|r|r|r|}
        \hline
        \multicolumn{1}{|c|}{\textbf{Metric}} & 
        \multicolumn{1}{c|}{\textbf{Mean}} & 
        \multicolumn{1}{c|}{\textbf{SD}} & 
        \multicolumn{1}{c|}{\textbf{Min}} & 
        \multicolumn{1}{c|}{\textbf{Max}} \\
        \hline
        \hline
        Repository Age (days) & 1555.02 & 885.56 & 244.00 & 5611.00 \\
        \hline
        Stars & 74.35 & 620.01 & 0.00 & 11963.00 \\
        \hline
        Forks & 11.24 & 99.99 & 0.00 & 1966.00 \\
        \hline
        Commits & 333.38 & 3046.94 & 1.00 & 65860.00 \\
        \hline
        Commits (6 months) & 5.04 & 48.24 & 0.00 & 954.00 \\
        \hline
        Contributors & 4.22 & 33.19 & 1.00 & 698.00 \\
        \hline
        Total Issues & 29.44 & 392.02 & 0.00 & 8363.00 \\
        \hline
        Open Issues & 10.63 & 133.92 & 0.00 & 2855.00 \\
        \hline
        Closed Issues & 20.09 & 261.42 & 0.00 & 5508.00 \\
        \hline
        Avg. Issue Close Time (days) & 7.58 & 71.89 & 0.00 & 1099.08 \\
        \hline
    \end{tabular}
\end{table}

\noindent
\textbf{Key insights for RQ$_2$:} 
The findings suggest that the open-source game ecosystem is characterized by skewed activity patterns: a small subset of projects sustain high levels of development and community engagement, whereas the majority remain inactive or minimally maintained. This distribution highlights the challenges of sustaining development and cultivating active communities in the domain, potentially due to lack of funding, or shifting developer priorities in hobbyist projects.

\subsection{RQ$_3$:~What languages, engines/frameworks, and build/CI tools do open-source games use, and how large/complex are they?}

We analyzed the technological landscape and asset composition using two sources: developer-reported tags on Itch.io and automated engine detection from repositories. We identified 75 unique engines and frameworks (Table~\ref{tab:engine_analysis}). Tag-based reports place Godot at 18.1\% and Unity at 16.1\%, while 41.8\% of projects lack an explicit engine tag, which may reflect custom engines or incomplete metadata.

Automated detection covered 491 games and identified an engine in 260 projects (52.9\%). These results confirm the tag-based pattern: Godot appears in 24.4\% of projects and Unity in 17.1\%. Engines remain undetected in 231 projects (47.0\%), which points to heterogeneous stacks that include custom engines and lesser-known frameworks.

Tool use extends beyond engines. Graphics pipelines show broad use of Blender (4.9\%), Aseprite (4.9\%), GIMP (4.3\%), and Krita (2.4\%). For audio, Audacity appears in 3.5\% of projects alongside other tools. This diversity indicates that developers assemble project-specific toolchains rather than rely on a single integrated pipeline.

\begin{table}[!htb]
    \caption{Game engine and framework distribution in open-source games.}
    \label{tab:engine_analysis}
    \centering
    \small
    \begin{tabular}{|l|c|c|}
        \hline
        \textbf{Technology} & \textbf{Tag-Based Analysis} & \textbf{Detection-Based Analysis} \\
        \hline
        \hline
        Godot & 89 (18.1\%) & 120 (24.4\%) \\
        \hline
        Unity & 79 (16.1\%) & 84 (17.1\%) \\
        \hline
        GameMaker & 13 (2.6\%) & 9 (1.8\%) \\
        \hline
        Unreal Engine & 2 (0.4\%) & 3 (0.6\%) \\
        \hline
        LÖVE & 3 (0.6\%) & 10 (2.0\%) \\
        \hline
        libGDX & 7 (1.4\%) & 7 (1.4\%) \\
        \hline
        Phaser & 5 (1.0\%) & 5 (1.0\%) \\
        \hline
    \end{tabular}
\end{table}

The asset analysis in Table~\ref{tab:asset_analysis} shows that images dominate project resources (mean 152.39; SD 950.50; max 20{,}196), indicating wide variation in scope and asset intensity. Audio assets average 22.98 files per project. 3D assets average 6.82, which aligns with the 2D focus of many projects. Shaders average 9.60, indicating the use of programmable visual effects even in primarily 2D games.

\begin{table}[!htb]
    \caption{Asset composition across open-source game projects.}
    \label{tab:asset_analysis}
    \centering
    \small
    \begin{tabular}{|l|r|r|r|r|}
        \hline
        \multicolumn{1}{|c|}{\textbf{Asset Type}} & 
        \multicolumn{1}{c|}{\textbf{Mean}} & 
        \multicolumn{1}{c|}{\textbf{SD}} & 
        \multicolumn{1}{c|}{\textbf{Min}} & 
        \multicolumn{1}{c|}{\textbf{Max}} \\
        \hline
        \hline
        Images & 152.39 & 950.50 & 
        0 & 20,196 \\
        \hline
        Audio & 22.98 & 101.36 & 
        0 & 1,802 \\
        \hline
        3D Models & 6.82 & 42.68 & 
        0 & 608 \\
        \hline
        Shaders & 9.60 & 134.00 & 
        0 & 2,964 \\
        \hline
        Fonts & 2.27 & 4.75 & 
        0 & 41 \\
        \hline
        Videos & 0.30 & 4.48 & 
        0 & 97 \\
        \hline
        Tilemaps & 2.25 & 28.58 & 
        0 & 543 \\
        \hline
    \end{tabular}
\end{table}

\noindent
\textbf{Key insights for RQ$_3$:}
Although there is a degree of technological heterogeneity, the vast majority of open-source game projects adopt either Godot or Unity as their engine, likely because Godot's open-source license and Unity's extensive asset ecosystem reduce barriers for indie developers. Asset composition indicates a predominant emphasis on 2D-oriented development.

\subsection{RQ$_4$: To what extent do open-source game projects document or adopt testing-related practices?}

We examined four facets: (i) presence of automated tests; (ii) information on bug reporting and fixing; (iii) documentation of testing and troubleshooting practices; and (iv) guidance for contributors on testing tasks. Across 491 projects, 36 include automated tests (7.33\%). Sixty-nine projects provide information on bug reporting or fixing (14.05\%). Forty document testing practices (8.15\%), and 11 reference troubleshooting activities (2.24\%). Twenty-one projects offer contributor guidance for testing (4.28\%). Overall, few projects expose test automation or formalize testing workflows, and projects rarely specify how contributors can participate in testing.

\noindent
\textbf{Key insights for RQ$_4$:}
Our findings suggest that testing practices are often informal, 
under-documented, and unevenly adopted across the game development landscape. This pattern reflects an ad hoc approach to software quality assurance within the open-source game domain, possibly prioritizing rapid prototyping and ``fun'' over rigorous testing due to the creative nature of game development.

\subsection{RQ$_5$: To what extent is the presence of test files associated with repository activity and user-perceived quality in open-source game projects?}

We tested whether repositories with test files show higher activity (commit counts) and higher user ratings. We estimated two regressions with the presence of test files as the main predictor and included repository age, stars, and forks as covariates to control for maturity and visibility.

For repository activity, we modeled commit counts with a Negative Binomial generalized linear model and a log link to handle overdispersed counts. Exponentiated coefficients yield rate ratios. We computed HC3 heteroskedasticity-consistent standard errors. The model is:

\begin{align}
\label{eq:nb_commits}
\log \big( \mathbb{E}[C_i] \big)
&= \beta_0
   + \beta_1\,T_i
   + \beta_2\,A_i
   + \beta_3\,S_i
   + \beta_4\,F_i
\end{align}

\noindent
where $C_i$ is the number of commits for repository $i$, $T_i$ indicates the presence of test files, and $A_i$, $S_i$, $F_i$ denote repository age, stars, and forks.

Table~\ref{tab:nb-commits} reports a statistically significant association between tests and activity. Holding age, stars, and forks constant, repositories with tests have an expected $3.34\times$ higher commit rate (95\% CI [1.98, 5.64]; $p<0.001$) than repositories without tests.


\begin{table}[!htb]
  \caption{NB GLM for commit counts (log link).}
  \label{tab:nb-commits}
  \centering
  \small
  \setlength{\tabcolsep}{4pt}
  \renewcommand{\arraystretch}{1.0}
  \begin{tabular*}{\columnwidth}{@{\extracolsep{\fill}} lccc c}
    \toprule
    \textbf{Predictor} & $\boldsymbol{\beta}$ & \textbf{Rate Ratio} & \textbf{95\% Confidence Interval} & \textbf{$p$} \\
    \midrule
    Intercept &  3.6746 & 39.432 & [26.128, 59.510] & <0.001 \\
    $T$       &  1.2056 &  3.339 & [ 1.976,  5.640] & <0.001 \\
    $A$       &  0.0007 &  1.001 & [ 1.000,  1.001] & <0.001 \\
    $S$       &  0.0013 &  1.001 & [ 0.998,  1.004] &  0.396 \\
    $F$       & -0.0019 &  0.998 & [ 0.991,  1.005] &  0.597 \\
    \bottomrule
  \end{tabular*}
  \\[2pt]
  \footnotesize Rate Ratio $=\exp(\beta)$. $C_i$: commits; $T$: tests present; $A$: repo age; $S$: stars; $F$: forks. 
\end{table}

The second model treats user ratings ($R_i$) as a continuous outcome and relates them to the presence of test files and covariates:

\begin{align}
\label{eq:ols_ratings}
R_i &= \beta_0
   + \beta_1\,T_i
   + \beta_2\,A_i
   + \beta_3\,S_i
   + \beta_4\,F_i
\end{align}

\noindent
where $R_i$ is the average rating for repository $i$; $T_i$ indicates the presence of test files; and $A_i$, $S_i$, $F_i$ denote repository age, stars, and forks.

Table~\ref{tab:ols-ratings} summarizes the OLS results. After controlling for $A$, $S$, and $F$, the coefficient on tests is negative and not statistically significant ($\beta=-0.09$, 95\% CI [$-0.39$, $0.21$]; $p=0.545$). Thus, within this sample, the presence of test files is not associated with higher (or lower) user ratings. None of the controls are significant, and the model explains little variance ($R^2=0.009$), indicating that repository-level characteristics considered here have limited power to account for rating differences.

\begin{table}[!htb]
  \caption{OLS regression for user ratings.}
  \label{tab:ols-ratings}
  \centering
  \small
  \setlength{\tabcolsep}{4pt}
  \renewcommand{\arraystretch}{1.0}
  \begin{tabular*}{\columnwidth}{@{\extracolsep{\fill}} lccc c}
    \toprule
    \textbf{Predictor} & $\boldsymbol{\beta}$ & \textbf{Std. Error} & \textbf{95\% Confidence Interval} & \textbf{$p$} \\
    \midrule
    Intercept &  4.5751 & 0.091 & [4.396, 4.754] & <0.001 \\
    $T$       & -0.0916 & 0.151 & [-0.388, 0.205] &  0.545 \\
    $A$       & -0.0000 & 0.000 & [-0.0001, 0.0001] &  0.581 \\
    $S$       &  0.0001 & 0.000 & [-0.0001, 0.0004] &  0.353 \\
    $F$       & -0.0005 & 0.001 & [-0.002, 0.001] &  0.589 \\
    \bottomrule
  \end{tabular*}
  \\[2pt]
  \footnotesize $R_i$: user rating; $T$: tests present; $A$: repo age; $S$: stars; $F$: forks. 
\end{table}

\noindent
\textbf{Key insights for RQ$_5$:}
Our findings suggest that although the presence of test files is linked to higher project activity, this increased activity does not appear to yield measurable improvements in player-perceived quality. Players may prioritize experiential factors like gameplay and narrative over underlying software stability, especially if bugs are non-critical, raising questions about the value of testing in user-facing metrics.

\section{Discussion}
\label{sec:discussion}

\paragraph{Who builds these games.}
Open-source games in our corpus align with what \citet{Mayerowitz2025} call ``Kei Games'': projects by individual developers or small teams that target focused scopes and short feedback loops. The genre mix (Puzzle, Action, Platformer), 2D-heavy asset inventories, and the dominance of Godot and Unity support this profile. Commercial titles such as \emph{Undertale}, \emph{Papers, Please}, and \emph{Terraria} illustrate the style, even though they are not open source.

\paragraph{Testing and sustainability.}
Only 36 projects (7.33\%) include automated tests, in line with prior work that reports low test adoption in games \cite{kasurinen_what_2014, PolitowskiEPetrillo2021}. Yet repositories with tests show a $3.34\times$ higher expected commit rate (RQ$_5$), controlling for age, stars, and forks. Tests may act as a proxy for an \textbf{engineering discipline}, or they may co-evolve with ongoing work. We cannot claim causality, but the association suggests that lightweight testing can support sustained maintenance.

\paragraph{Why tests do not move ratings.}
We find no association between the presence of tests and user ratings. Several factors can explain this result: (i) ratings are sparse (58.9\% of projects have no ratings), which limits signal; (ii) players reward content, feel, and novelty more than backend quality; and (iii) ratings have ceiling effects (means above 4.0) that compress variance. Repository-level controls (age, stars, forks) also do not explain ratings, and the OLS model has low $R^2$. Quality as perceived by players likely depends on factors outside our metadata (moment-to-moment design, pacing, art direction, community, and platform visibility). This lack of correlation prompts a hypothesis: Do players care about software quality? Our data suggests not directly, as ratings may capture ``fun'' and engagement rather than technical robustness. However, metadata alone cannot fully answer this.

\paragraph{Documentation and contributor pathways.}
Projects rarely document testing or guide contributors: 14.05\% mention bug reporting/fixing, 8.15\% document testing, 2.24\% mention troubleshooting, and 4.28\% provide testing guidance. This limits outside help and slows defect resolution. Simple steps can raise the floor: add \texttt{CONTRIBUTING.md} with reproduction steps and platforms, template issue forms for bugs, a short ``how to run tests'' section, and a crash/diagnostics checklist.

\paragraph{Implications for practice.}
Low test adoption and sparse contributor guidance suggest that many projects lack a feasible on-ramp for quality work. Teams can introduce a small test core that fits tight schedules and common engines. Run scene-load smoke tests headless through engine CLIs to catch broken builds early. Record and replay input traces to detect gameplay regressions without bespoke harnesses. Package these steps in one GitHub Actions workflow so contributors can build and run checks on every pull request. These measures raise confidence, reduce back-and-forth on bug reports, and make outside contributions safer to merge.

\paragraph{Implications for research.}
Results point to concrete tooling and study designs. Build engine-specific runners for Godot and Unity that expose assertions, timing, and deterministic replay without editor dependencies. Add asset-aware regression checks that diff tilemaps, sprites, and shader outputs to catch visual changes that code coverage misses. Provide privacy-preserving crash bundles and repro scripts that small teams can attach to issues. Define catalogues of engine-level smells and auto-fixes, and evaluate their effect on build stability and review time. Run longitudinal studies that track projects before and after test introduction to probe causality behind the observed activity lift, using outcomes such as survival, release cadence, mean time to close, and regression rates.

\paragraph{What the dataset adds.}
The dataset contributes a metadata view that joins genres, activity, engines, assets, and testing signals. It supports replication and cross-study comparisons without requiring source distribution. It also exposes gaps: missing ratings, incomplete engine tags, and silent issue trackers. Future releases can link runtime artifacts (crash dumps, CI logs) and release histories to enrich analyses.

\paragraph{Limitations and scope.}
Our sampling of Itch.io and GitHub introduces selection bias toward public and smaller projects. However, this focus is deliberate and valuable, as itch.io supports a vibrant open-source indie community, enabling access to artifacts opaque in commercial settings and aligning with calls for more empirical data on small-team practices \cite{engstrom_game_2018}. Engine detection can miss custom or atypical layouts, potentially undercounting diverse stacks; we addressed this flagging ambiguous cases for manual review. Genre is multi-label and noisy. Stars, forks, and ratings are coarse proxies for impact and quality. We mitigate these risks with manual checks, including cross-validation by multiple researchers as detailed in Section 4. These limits do not alter the main signals: most projects are small, 2D-focused, and engine-driven; tests are rare yet associated with higher activity; ratings, when present, do not track test presence.

\section{Concluding Remarks}
\label{sec:concluding_remarks}

This paper presented an empirical analysis of 491 open-source games revealing patterns in development: focused genres like Puzzle, Action, and Platformer; engine reliance on Godot (24.4\%) and Unity (17.1\%); 2D-centric, image-heavy assets; and ad hoc testing (only 7.33\% automated) that correlates with higher activity (3.34× commit rate) but not perceived quality. These align with ``Kei Games''\cite{Mayerowitz2025} by small teams prioritizing creativity over rigor, confirming survey evidence with code-level data.

The findings also provide supporting evidence that Godot and Unity are the most widely adopted engines. 
In addition, we found that 
testing practices remain underdeveloped across the ecosystem. 
Although projects that include test files exhibit higher levels of repository activity, 
the presence of such files does not appear to yield measurable improvements in player-perceived quality.

Limitations include itch.io's bias and rating sparsity, potentially understating commercial parallels. As future work, we intend to broaden both the scope and depth of our investigation. 
This includes extending data collection to additional platforms and expanding the sample of open-source games, as well as conducting qualitative studies of projects that have success both in terms of their open-source sustainability and their popularity on platforms such as itch.io and Steam. Such studies could build on our dataset to test hypotheses like the role of testing in long-term project health or the impact of engine choice on developer retention.

\begin{acks}

This work is supported by National Council for
Scientific and Technological Development – CNPq (grants nr. 444311/2024-6 and 406941/2025-4).
H.M. Canella is financially supported by CNPq (grant nr. 141714/2025-5).

\end{acks}

\balance
\bibliographystyle{ACM-Reference-Format}
\bibliography{sample-base}
\end{document}